\documentclass[journal]{IEEEtran}

\usepackage{cite}

\usepackage{graphicx}

\usepackage{amsmath}

\usepackage{array}

\usepackage{fixltx2e}

\usepackage{url}

\begin{document}

\title{An analog ac voltage amplifier based on a single straintronic magnetic tunnel junction}
%
%
%

\author{Cael Johnson~\IEEEmembership{Student~Member,~IEEE}, Rahnuman Rahman~\IEEEmembership{Student~Member,~IEEE} and Supriyo~Bandyopadhyay,~\IEEEmembership{Life~Fellow,~IEEE} \\

\thanks{Manuscript received ...; revised ... 
 (\it Corresponding author: Supriyo Bandyopadhyay)}
\thanks{The authors are with the Department
of Electrical and Computer Engineering, Virginia Commonwealth University, Richmond, VA, 23284 USA. (email:  sbandy@vcu.edu). Cael Johnson was supported by a NSF REU grant DMR-2349694.}} 

\maketitle

\begin{abstract}
 Magnetic tunnel junctions (MTJs) are known for their digital applications (memory and logic).  A special class of them called ``straintronic'' magnetic tunnel junctions (s-MTJ) has lately emerged as a potential platform for analog applications because their conductance can be varied {\it continuously} with mechanical strain generated with a gate voltage. The  conductance versus gate voltage (transfer) characteristic {\it always} has a linear region and that can be leveraged for a variety of analog applications. Here, we discuss one such application, namely analog voltage amplification. If the s-MTJ's gate voltage is fixed around the midpoint of the linear region and a small ac voltage is superimposed on it, then the ac voltage can be amplified without distortion as long as its amplitude is small enough to avoid gate voltage excursion beyond the linear region. Unlike a transistor-based voltage amplifier whose amplification is determined solely by the transistor's internal parameters -- namely the transconductance and Early resistance -- here the amplification can be varied by an external power supply voltage. We examine the maximum allowed amplitude and frequency of input signal for distortion-free amplification by modeling the magnetization dynamics and derive an expression for the amplification.
\end{abstract}


%
\IEEEpeerreviewmaketitle

\section{Introduction}

\IEEEPARstart{T}{he} ``straintronic magnetic tunnel junction'' (s-MTJ) is a special type of magnetic tunnel junction
(MTJ) whose conductance can be changed continuously or gradually from high
to low, or vice versa, with a gate voltage that generates strain in the s-MTJ's
magnetostrictive soft layer \cite{d'souza,APR,monograph,npj}. This unusual feature of continuous variation, not available in MTJs
that are switched abruptly with spin transfer torque, spin–orbit torque or
voltage-controlled-magnetic anisotropy, enables many {\it analog} applications where the typically low tunneling
magneto-resistance ratio of MTJs is not a serious impediment unlike in digital
logic or memory. 

Interestingly, the conductance versus gate voltage (transfer) characteristic of a s-MTJ always exhibits a {\it linear region} that can be
exploited to implement analog arithmetic, vector matrix multiplication and linear
synapses in deep learning networks \cite{jphysd}. The existence of this linear region was proved analytically in ref. \cite{rahnuma}. It is an exceptional feature since it is not found in most other electron devices such as MOSFETs or memristors. Here, we present a special application of this unique feature, namely the implementation of a novel analog ac voltage amplifier.

\begin{figure}[!t]
\centering
\includegraphics[width=0.46\textwidth]{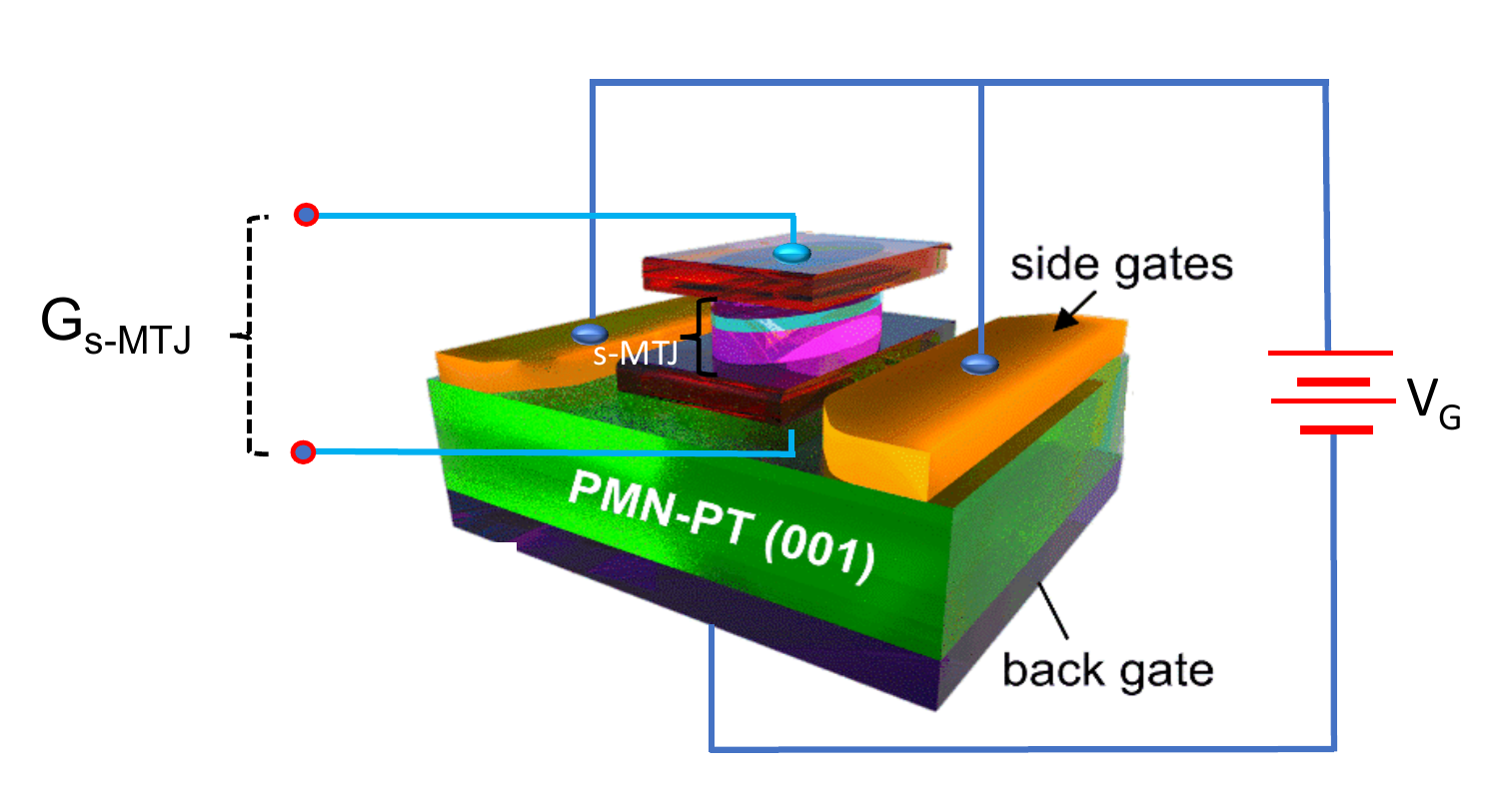}
\caption{The structure of a straintronic magnetic tunnel junction. Reprinted from \cite{zhao}, with the
permission of AIP Publishing.}
\label{fig:zhao}
\end{figure}

The operation of a s-MTJ has been demonstrated by many groups \cite{li,zhao,chen,chen2}. The configuration used in \cite{zhao} is shown in Fig. \ref{fig:zhao} and consists of a MTJ delineated on a piezoelectric thin film or substrate, such as (001) PMN-PT. The piezoelectric is poled in the vertically up direction. A gate voltage $V_G$ is applied between the
two side gates (shorted together) and the back gate. Initially, the magnetizations of the hard and soft layers of the MTJ are mutually antiparallel because of any residual dipole interaction between the two. The strain generated in the piezoelectric by the gate voltage  rotates the
magnetization of the soft layer
through an angle $\phi$ and that changes the resistance $R(\phi)$ of the s-MTJ according to the
equation $R(\phi) = R_P + \left ( R_{AP} - R_P \right )/2 \left [ 1 + cos \phi \right]$, where $R_P$ is the MTJ's resistance when the magnetizations of the hard and soft layer are parallel and $R_{AP}$ is the resistance when they are antiparallel. 

Ref. \cite{rahnuma} showed analytically that over a
range of gate voltage $V_G$, the transfer characteristic (s-MTJ conductance $G_{s-MTJ}$ versus gate voltage $V_G$) will be linear and obey the relation
\begin{equation}
    G_{s-MTJ} = 1/R_{AP} + \kappa \left (V_G - \delta \right ),
    \label{eq:linear}
\end{equation}
where $\kappa = -1/\left ( 2 R_{AP} \Gamma \right )$, $\delta = \Gamma - \nu$, $\Gamma = \mu_0M_s |H_d| t/\left ( 3 \lambda_s Y d_{33} \right )$, $\nu = \left [ M_s \left (N_{min} - N_{maj} \right )/|H_d| \right ] \Gamma$. Here $\mu_0$ is the permeability of free space, $M_s$
is the saturation magnetization of
the soft layer, $t$ is the thickness of the piezoelectric layer, $\lambda_s$
is the saturation magnetostriction of the soft layer, $Y$ is the Young’s modulus of the soft layer, $d_{33}$
is the diagonal component of the piezoelectric coefficient tensor of the
piezoelectric layer, $H_d$ is the effective magnetic field seen by the soft layer due to dipole interaction with the hard layer, while $N_{min}$ and $N_{maj}$ are the demagnetization factors along the
minor and major axis of the elliptical soft layer, which depend on the dimensions
of the soft layer \cite{chikazumi}. The linear relation in Eq. (\ref{eq:linear}) was verified by
stochastic Landau–Lifshitz-Gilbert-Langevin (LLGL) simulations at room temperature in \cite{rahnuma}.

\begin{figure}[!t]
\centering
\includegraphics[width=0.46\textwidth]{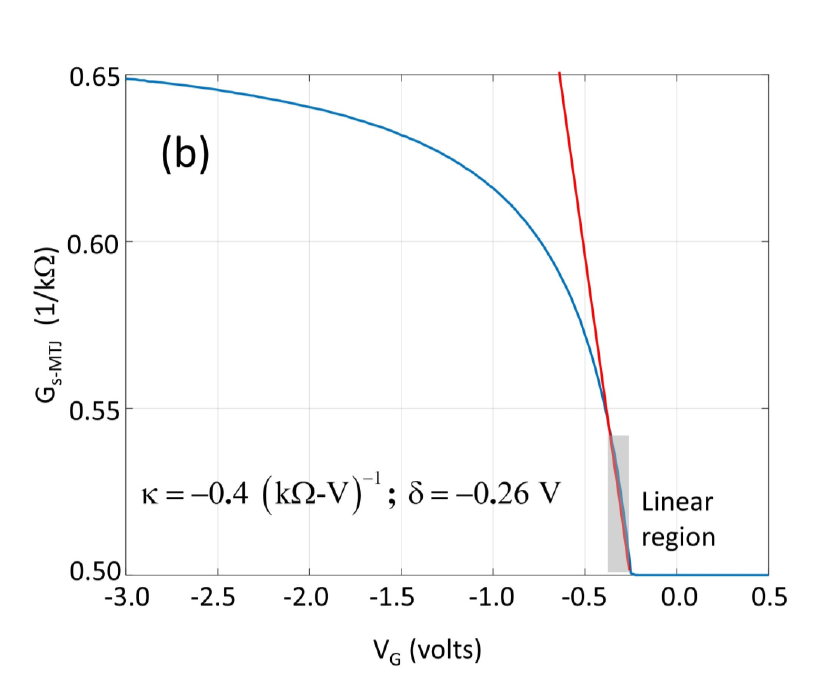}
\caption{The computed $G_{s-MTJ}$ versus $V_G$ transfer characteristic of a s-MTJ at room temperature. The linear region is shaded in gray. Within this region, the characteristic follows Equation (\ref{eq:linear}). The values of $\kappa$ and $\delta$ for this MTJ are shown. \copyright 2022 IEEE. Reprinted with permission from \cite{rahnuma}.}
\label{fig:charac}
\end{figure}

\begin{figure}[!b]
    \centering
    \includegraphics[width=0.99\linewidth]{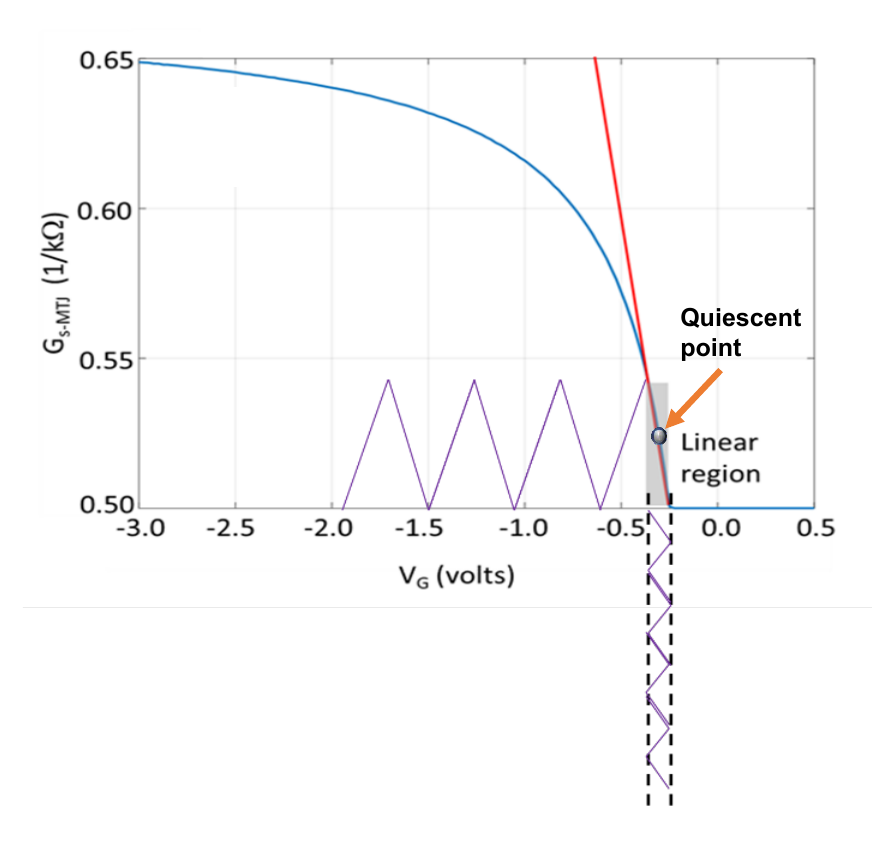}
    \caption{A swing of the gate voltage around the quiescent point results in a corresponding swing of the s-MTJ conductance. The waveform shown here is triangular but the principle holds for any arbitrary waveform. The gate voltage swing has to be small enough to stay within the linear region.}
    \label{fig:swing}
\end{figure}

\begin{figure*}[!hbt]
    \centering
    \includegraphics[width=0.99\linewidth]{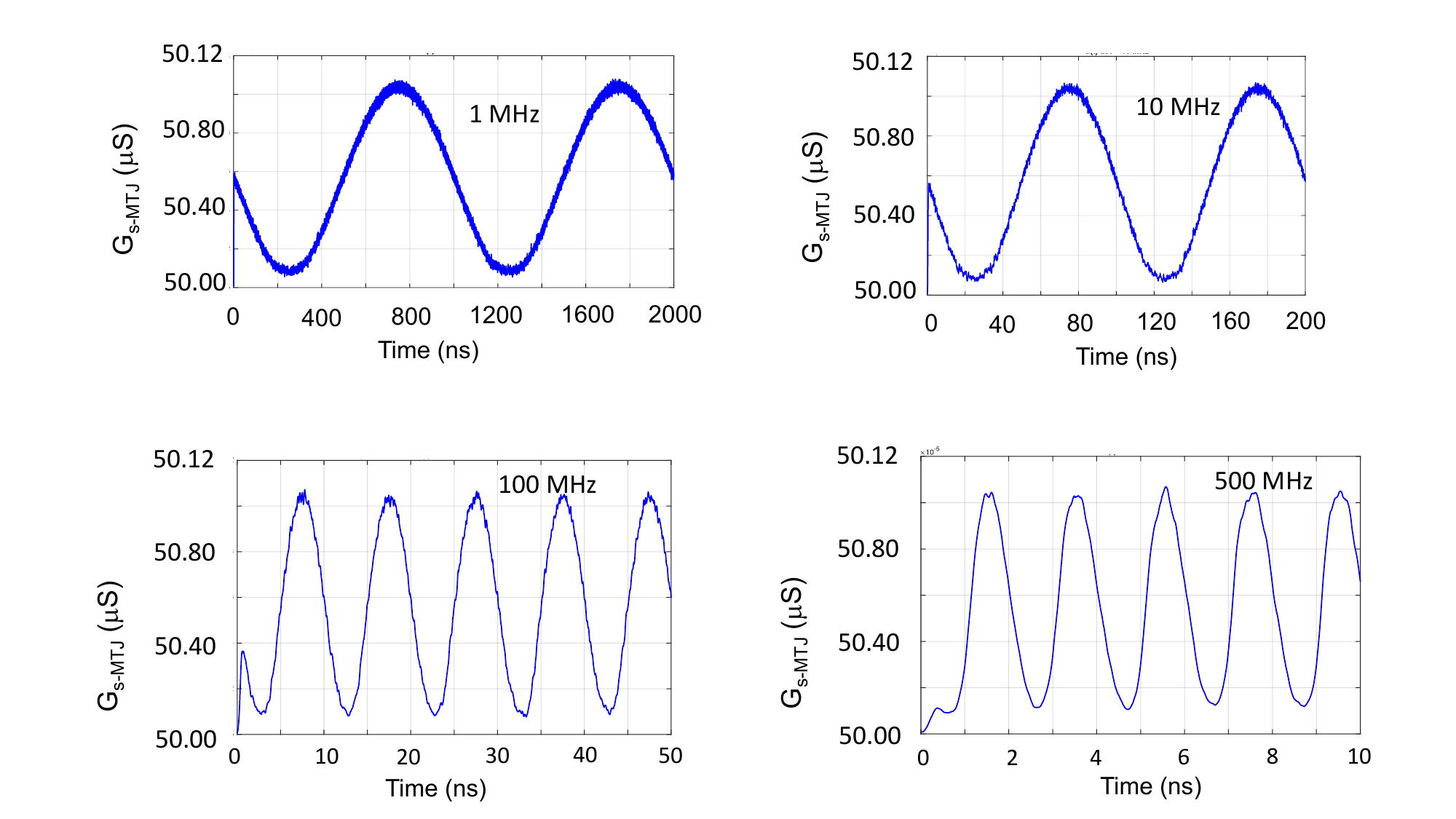}
    \caption{Time variation of the s-MTJ conductance in the presence of a dc + ac gate voltage at different frequencies. The dc voltage was -0.26 V and the ac voltage amplitude was 10 mV to ensure that the gate voltage excursion always remains within the linear region of the transfer characteristic. }
    \label{fig:results}
\end{figure*}
\begin{figure}[!hbt]
    \centering
    \includegraphics[width=0.99\linewidth]{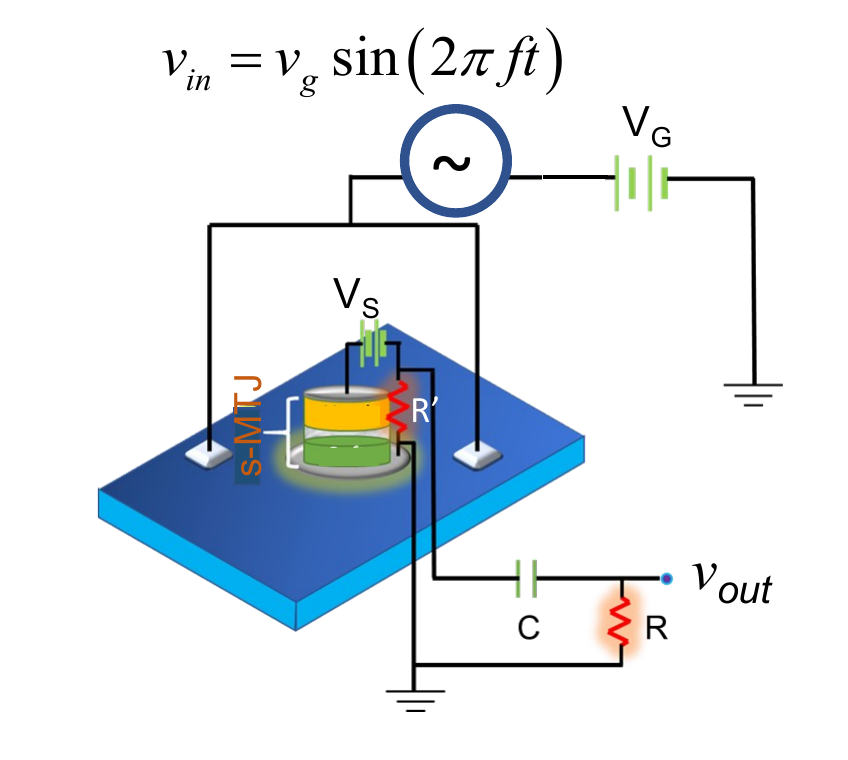}
    \caption{The ac amplifier circuit realized with a single s-MTJ.}
    \label{fig:amp}
\end{figure}

Ref. \cite{rahnuma} also used room-temperature LLGL equations to obtain the $G_{s-MTJ}$ versus $V_G$ relation for a s-MTJ of elliptical cross section with the following soft layer parameters: major axis = 800 nm, minor axis = 700 nm, thickness = 2.2 nm, $M_s = 8 \times 10^5$ A/m, $H_d$ = 1000 Oe, Gilbert damping constant = 0.1, $\lambda_s$ = 600 ppm, $Y$ = 120 GPa, $d_{33}$ = 1.5$\times$10$^{-9}$ C/N, $t$ = 1 $\mu$m, $R_{AP}$ = 2 k$\Omega$ and $R_P$ = 1 k$\Omega$. The computed characteristic is shown in Fig. \ref{fig:charac}. 

\section{Results and Discussion}

If we bias the s-MTJ with a dc gate voltage at the midpoint of the linear region and then superimpose an alternating voltage on it, the s-MTJ conductance $G_{s-MTJ}$ will swing around the quiescent point as shown in Fig. \ref{fig:swing}.  The amplitude of the alternating voltage has to be small enough to ensure that the gate voltage excursion does not take it outside the linear region. With appropriate circuitry, the conductance swing can be converted to a voltage swing which is an amplified version of the superimposed alternating voltage. This implements a voltage amplifier.

We repeated the LLGL simulations of ref. \cite{rahnuma} by fixing the $V_G$ value at -0.26 V (approximately the midpoint of the linear region in Fig. \ref{fig:charac}) and superimposed a sinusoidal ac voltage of 10 mV amplitude and varying frequency. This amplitude ensures that the gate voltage swing does not go past the linear region. The simulation details are the same as in ref. \cite{rahnuma} and not repeated here to avoid redundancy. We then obtained the time-varying conductance (conductance swing) at room temperature. These results are shown in Fig. \ref{fig:results}

We found that the maximum ac gate voltage amplitude allowed in this case is 16 mV. Beyond that, the gate voltage excursion goes beyond the linear region in the transfer characteristic and the waveform begins to develop a half-wave rectified (or clipped) shape since the slope $\partial G_{s-MTJ}/\partial V_G = \kappa$  no longer remains constant.

It is clear from Fig. \ref{fig:results} that there is a cut-off frequency beyond which the conductance oscillation no longer follows the ac gate voltage oscillation's sinusoidal shape, and distortion begins to appear. This happens because the magnetization dynamics can no longer follow the gate voltage dynamics quasi-statically when the time scale of the evolution of magnetization (related to the magnetization precession period) becomes longer than the gate voltage period. Ferromagnets typically have ferromagnetic resonance frequencies of a few GHz and once the signal frequency begins to approach that limit, the magnetization temporal evolution can no longer keep pace with the signal frequency. That is the reason why the conductance oscillation gets distorted from the sinusoidal shape above 100 MHz frequency.

The above results can be leveraged to implement an ac analog voltage amplifier which is shown in Fig. \ref{fig:amp} where $V_s$ is a dc supply voltage.

In Fig. \ref{fig:amp}, the voltage over the resistor $R'$ is 
\begin{equation}
    V = \frac{R'}{R' + R_{s-MTJ}} V_s \approx \frac{R'}{R_{s-MTJ}} V_s = G_{s-MTJ}R'V_s.
\end{equation}
if $R_{s-MTJ} \gg R'$.

The blocking capacitor $C$ filters out any dc offset and hence the voltage over the resistor $R$ is 
\begin{equation}
    \tilde{V} = {\tilde{G}_{s-MTJ}}\left (R \parallel R' \right ) V_s,
\end{equation}
where $\tilde{V}$ is the filtered ac voltage over the load resistor $R$ and ${\tilde{G}_{s-MTJ}}$ is the ac component of the conductance oscillations which is given by
\begin{equation}
    {\tilde{G}_{s-MTJ}} = \frac{\partial G_{s-MTJ}}{\partial V_G}v_g sin (2 \pi f t) = \kappa v_g sin (2 \pi f t),
\end{equation}
where $v_g$ is the amplitude of the ac component of the gate voltage and $f$ is the frequency. Since in the linear region of the transfer characteristic, the slope $\partial G_{s-MTJ}/\partial V_G = \kappa$ is a constant, we get that 
\begin{equation}
    \tilde{V} = \kappa v_g sin (2 \pi f t) \left (R \parallel R' \right ) V_s,
\end{equation}
which yields a linear amplification of $A = \tilde{V}/v_g$ = $\kappa \left (R' \parallel R \right ) V_s$.

\section{Conclusion}

To our knowledge, MTJs have not been used for analog amplification in this fashion. Analog applications of MTJs are few and far between. Here, we draw attention to the fact that straintronic MTJs can have an intriguing application as an analog voltage amplifier. The operating frequency of the amplifier is limited by the speed of the magnetization dynamics in ferromagnets and it may be possible to increase that by replacing ferromagnets with compensated or uncompensated ferrimagnets because they are known to have faster dynamics \cite{liu}. 

The allowable input voltage amplitude can be increased if we can increase the extent of the linear region by decreasing the slope $\partial G_{s-MTJ}/\partial V_G = \kappa$. This can be done by increasing the s-MTJ antiparallel resistance $R_{AP}$ by decreasing the cross-sectional area and/or increasing the thickness of the spacer region.

Finally, one difference with traditional transistor amplifiers is that in the latter the amplification is determined solely by internal parameters, namely the transconductance and Early resistance \cite{millman}. Here, the amplification is proportional to the supply voltage $V_s$ and hence can be increased at will by increasing $V_s$.

\pagebreak


\begin{thebibliography}{1}

\bibitem{d'souza}
N. D'Souza, et al., ``Energy-efficient switching of nanomagnets for computing: Straintronics and other methodologies'', { \it Nanotechnology}, vol. 29, No. 44, 442001 (2018). https://doi.org/10.1088/1361-6528/aad65d.
\bibitem{APR}
S. Bandyopadhyay, J. Atulasimha and A. Barman, ``Magnetic straintronics: Manipulating the magnetization of magnetostrictive nanomagnets with strain for energy-efficient applications'', {\it Appl. Phys. Rev.}, vol. 8, 041323 (2021). https://doi.org/10.1063/5.0062993.
\bibitem{monograph}
S. Bandyopadhyay, {\it Magnetic Straintronics:
An Energy-Efficient Hardware Paradigm for Digital and Analog Information Processing} (Springer-Nature, Cham, Switzerland, 2022). DOI:10.1007/978-3-031-20683-2.
\bibitem{npj}
S. Bandyopadhyay, ``Perspective: There is plenty of room for magnetic straintronics in the analog domain'', {\it npj Spintronics}, vol. 2, 15 (2022). https://doi.org/10.1038/s44306-024-00018-3.
\bibitem{jphysd}
S. Bandyopadhyay, ``Straintronic magnetic tunnel junctions for analog computation: a perspective'', {\it J. Phys. D: Appl. Phys.}, vol. 58, 152001 (2025). DOI 10.1088/1361-6463/adb9f7.
\bibitem{rahnuma}
R. Rahman and S. Bandyopadhyay, ``A nonvolatile all-spin nonbinary matrix
multiplier: An efficient hardware
accelerator for machine learning'', {\it IEEE Trans. Elec. Dev.}, vol. 69, 7120-7127 (2022).
 DOI: 10.1109/TED.2022.3214167.
\bibitem{li}
P. Li, A. Chen, D. Li, Y. Zhao, S. Zhang, L. Yang, Y. Liu, M. Zhu, H. Zhang, and X. Han, {\it Adv. Mater.}, vol. 26, 4320 (2014). https://doi.org/10.1002/adma.201400617
\bibitem{zhao}
Z. Zhao, M. Jamali, N. D’Souza, D. Zhang, S. Bandyopadhyay, J. Atulasimha and J-P Wang, 
``Giant voltage manipulation of MgO-based magnetic tunnel junctions via localized
anisotropic strain: a potential pathway to ultra-energy-efficient memory technology'', {\it Appl.
Phys. Lett.}, vol. 109, 092403 (2016). https://doi.org/10.1063/1.4961670.
\bibitem{chen}
A. Chen, Y. Wen, B. Fang, Y. Zhao, Q. Zhang, Y. Chang, P. Li, H. Wu, H. Huang, Y. Lu, Z. Zeng, J. Cai, X. Han, T. Wu, X. X Zhang and  Y. Zhao, ``Giant nonvolatile manipulation of magnetoresistance in magnetic tunnel junctions by electric fields via magnetoelectric coupling'', {\it Nat. Commun.}, vol. 10, 243 (2019). https://doi.org/10.1038/s41467-018-08061-5.
\bibitem{chen2}
A. Chen, Y. Zhao, Y. Wen, L. Pan, P. Li and X. X. Zhang, ``Full voltage manipulation of the resistance of a magnetic tunnel junction'', {\it Sci. Adv.}, vol. 5, eaay5141 (2019). DOI: 10.1126/sciadv.aay5141.
\bibitem{chikazumi}
S. Chikazumi, {\it Physics of Magnetism} (Wiley, New York, 1964).
\bibitem{liu}
J. Finley and L. Liu, ``Spintronics with compensated ferrimagnets'', {\it Appl. Phys. Lett.}, vol. 116, 110501 (2020). https://doi.org/10.1063/1.5144076.
\bibitem{millman}
J. Millman and C. C. Halkias, {\it Electronic Devices and Circuits} (Tata-McGraw Hill, New Delhi, 1999).


\end{thebibliography}
\end{document}